\documentclass{emulateapj}
\usepackage{apjfonts}
\usepackage{color}
\newcommand{\thetae}{\theta_{\rm E}}

\definecolor{darkbrown}{RGB}{139,69,19}

\shorttitle{C-R LENSING CHANNEL}
\shortauthors{HAN, SHIN, \& JUNG}
\begin{document}

\title{Detections of Planets in Binaries Through the Channel of Chang-Refsdal Gravitational Lensing Events}

\author{
Cheongho Han$^{1}$, In-Gu Shin$^{2}$, Youn Kil Jung$^{2}$  
}

% ===========================================================  
\affil{$^{1}$  Department of Physics, Chungbuk National University, Cheongju 361-763, Republic of Korea}
\affil{$^{2}$ Harvard-Smithsonian Center for Astrophysics, 60 Garden St., Cambridge, MA, 02138, USA}
% ===========================================================  

\begin{abstract}
Chang-Refsdal (C-R) lensing, which refers to the gravitational lensing of a point mass perturbed by 
a constant external shear, provides a good approximation in describing lensing behaviors of either a 
very wide or a very close binary lens. C-R lensing events, which are identified by short-term anomalies 
near the peak of a high-magnification lensing light curves, are routinely detected from lensing surveys, 
but not much attention is paid to them. In this paper, we point out that C-R lensing events provide an 
important channel to detect planets in binaries, both in close and wide binary systems. Detecting 
planets through the C-R lensing event channel is possible because the planet-induced perturbation occurs 
in the same region of the C-R lensing-induced anomaly and thus the existence of the planet can be 
identified by the additional deviation in the central perturbation. By presenting the analysis of 
the actually observed C-R lensing event OGLE-2015-BLG-1319, we demonstrate that dense and high-precision 
coverage of a C-R lensing-induced perturbation can provide a strong constraint on the existence of a 
planet in the wide range of the planet parameters. The sample of an increased number of microlensing 
planets in binary systems will provide important observational constraints in giving shape to the 
details of the planet formation scenario which has been restricted to the case of single stars.
\end{abstract}


\section{Introduction}

Majority of stars reside in binary systems \citep{Abt1983, Raghavan2010}. The mechanism of planet 
formation around binary systems would be different from that around single stars not only 
because the environment of the protoplanetary disk would be affected by the binary companion but 
also because the binary companion would affect the long-term stability of the planet orbit. Therefore, 
one of the most generic environments to be considered in the study of planet formation should be that 
of a binary. However, the major planet-formation scenarios that have been developed over the past 
decades, e.g., core-accretion theory \citep{Safronov1969, Goldreich1973, Hayashi1985, Pollack1996}
and disk instability theory \citep{Kuiper1951, Cameron1978, Boss2012}, were mostly focused on the 
case of single stars.

Although a considerable work has been done about the effect of binary companions on the planet 
formation \citep[e.g.,][]{Thebault2015} and the long-term orbital stability \citep[e.g.,][]{Szebehely1980}, 
this work was restricted to mostly theoretical studies and thus many details about the planet formation 
scenario remain uncertain. These details can be refined by the constraints provided by the sample of 
actually detected planetary systems. Unfortunately, there exists only a total of 19 known planets in 
17 binary systems and most of these planets exist under a similar environment, i.e. circumbinary 
planets orbiting very close binaries.  To give details about the planet formation in binary systems, 
therefore, it is important to detect more of such planets residing under various environments.

When a gravitational microlensing event is caused by a very wide binary object, the lensing 
behavior in the region 
around each lens component is approximated by Chang-Refsdal (C-R) lensing, which refers 
to the gravitational lensing of a point mass perturbed by a constant external shear $\gamma$ 
\citep{Chang1979, Chang1984}. In the low shear regime ($\gamma < 1$), C-R lensing induces a small astroid-shape 
caustic around the lens. The lensing behavior of a very close binary, on the other hand, can be 
approximated by a point-mass plus quadrupole lensing. In this case, an astroidal caustic similar 
to the C-R lensing caustic is produced around the center of mass of the binary. For this reason, 
the event caused by a very wide or a very close binary lens is often referred to as the C-R lensing 
event. Due to the existence of the caustic in the central region, C-R lensing events are identified 
by a short-term anomaly that appears near the peak of a very high-magnification event. Although C-R 
lensing events are routinely detected from microlensing surveys, not much attention is paid to them 
because they are thought to be simply one type of numerous binary-lens events and thus of little 
scientific importance.

In this work, we point out that C-R lensing events provide an important channel to detect planets 
in binaries, both in close and wide binary systems. In order to demonstrate that dense and 
high-precision coverage of a C-R lensing-induced perturbation can provide a strong constraint 
on the existence of a planet in the wide range of the planet parameters, we present analysis of 
the actually observed C-R lensing event OGLE-2015-BLG-1319.

The paper is organized as follows. In section 2, we briefly describe the lensing properties  
in the cases 
where the lens is composed of a single, binary, and triple masses. We also describe the C-R lensing 
behavior. In section 3, we estimate the detection efficiency of planets in binaries by conducting 
analysis of the lensing event OGLE-2015-BLG-1319.  We summarize the results and conclude in section 4.

\section{Chang-Refsdal Lensing Channel}

When a point-mass lensing event occurs the lensing behavior is described by the lens equation
\begin{equation}
\zeta = z - {1\over \bar{z}},
\label{eq1}
\end{equation}
where $\zeta=\xi + i\eta$ and $z=x+iy$ denote the complex notations of the source and image positions, 
respectively, and $\bar{z}$ represents the complex conjugate of $z$. Here all lengths are normalized 
to the angular Einstein ring radius $\thetae$ and the lens is positioned at the origin. Solving the 
lens equation yields two solutions of image positions: one outside and the other inside the Einstein 
ring. The magnification $A_j$ of each image $j$ is given by
\begin{equation}
A_j={1\over {\rm det} J_j}; \qquad
{\rm det} J_j=\left\vert 1-{\partial\zeta\over \partial\bar{z}}
{\partial\bar{\zeta}\over \partial z }
\right\vert_{z=z_j},
\label{eq2}
\end{equation}
where $J_j$ is the Jacobian of the lens equation evaluated at the image positions $z_j$ and 
${\rm det}J_j$ is the determinant of the Jacobian.  Since the individual microlensing images 
cannot be resolved, the observed lensing magnification is the sum of the magnifications of the 
individual images, i.e.\ $A=\sum_i A_i$. For a point mass, the magnification is represented 
analytically by
\begin{equation}
A=
{\vert \zeta\vert^2+2 \over \vert \zeta\vert  \sqrt{\vert \zeta\vert ^2+4}},
\label{eq3}
\end{equation}
where $\vert \zeta\vert =(t-t_0)/t_{\rm E} + i u_0  $ is the lens-source separation 
with the length normalized to $\thetae$, 
$t_{\rm E}$ is the Einstein time scale, $t_0$ is the time of the closest lens-source approach, 
and $u_0$ is the lens-source separation at $t_0$.  For a rectilinear relative lens-source motion, 
the lensing light curve is characterized by a smooth and symmetric shape \citep{Paczynski1986}.

When an event is produced by a lens composed multiple components, the lens equation is expressed as
\begin{equation}
\zeta = z - \sum_{i=1}^{N}{ \epsilon_i\over \bar{z}-\bar{z}_{L,i}},
\label{eq4}
\end{equation}
where $z_{L,i}$ and $\epsilon_i=m_i/m_{\rm tot}$, and $m_i$ represent the location, mass fraction, 
and mass of each lens component, respectively. The notation $N$ denotes the number of the lens 
components, and thus $N=2$ for a binary lens.  Here all lengths ($\zeta$ and $z$'s) are in units 
of the angular Einstein radius corresponding to the total mass of the lens, $\theta_{\rm E}$.  
One of the most important properties of a binary lens that differentiate from those of a single 
lens is the formation of caustics, which represent the sets of source positions at which 
${\rm det} J=0$ and thus the magnification of a point source becomes infinite.  As a result, 
a binary-lensing light curve can exhibit strong deviations when a source approaches close to 
or passes over the caustic.  Caustics of a binary lens form a single or multiple sets of closed 
curves and each curve composed of concave curves that meet at cusps. The topology of the 
binary-lens caustic is broadly classified into 3 categories \citep{Erdl1993, Danek2015}.  In 
the case of a binary where the binary separation is greater than 
$(\sqrt[3]{\epsilon_1} +\sqrt[3]{\epsilon_2})^{3/2}$ (wide binary), there exist two sets of 
4-cusp caustics that are located close to the individual lens components.  When the binary 
separation is smaller than $(\sqrt[3]{\epsilon_1} +\sqrt[3]{\epsilon_2})^{-3/4}$ (close binary), 
the caustic is composed of 3 pieces, where the central caustic with 4 cusps is formed around the 
center of mass of the binary lens, and the other 2 triangular caustics are located away from the 
center of mass.  In the intermediate separation region, there exists a single big caustic with 
6 cusps.  For the visual presentation of the binary caustic topology, see Figure 3 of 
\citet{Dominik1999}. In the extreme case where the binary separation is much greater or less 
than $\thetae$, the 4-cusp central caustic has an astroid shape which is symmetric with respect 
to the binary axis and the line vertical to the binary axis.

For the exact description of the lensing behavior produced by a lens system of a planet orbiting 
a binary object, one needs the triple lens equation, i.e.\ Eq.~(\ref{eq4}) with $N=3$. 
With the addition of a third lens component, the complexity of the lensing 
behavior greatly increases and caustics can exhibit 
self-intersection and nesting \citep{Rhie1997, Gaudi1998, Danek2015}.

% Figure 1 ------------------------------------------------------
\begin{figure}
\includegraphics[width=\columnwidth]{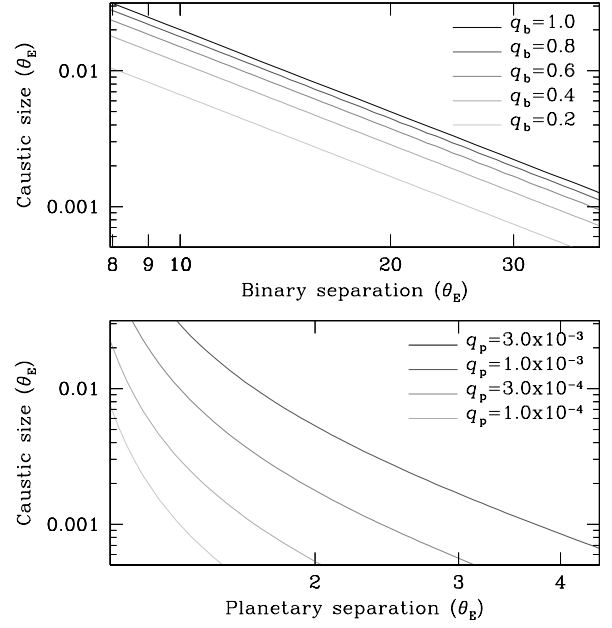}
\caption{Sizes of the C-R lensing caustic (upper panel) and the planet-induced central caustic 
as a function of the separation between the lens components.  Curves with different grey tone 
show the variation of the caustic size depending on the mass ratio between binary lens components 
$q_b$ (upper panel) and the mass ratio between the planet and the host $q_p$ (lower panel).  
Both the caustic size and the binary separation are normalized to the angular Einstein radius 
corresponding to the total lens mass.
}
\label{fig:one}
\end{figure}
% --------------------------------------------------------------

A planetary lens corresponds to the extreme case of a binary lens where the mass of one of the 
lens components is much smaller than the other, i.e.\ $\epsilon_1\sim 1$ and $\epsilon_2\ll 1$. 
In this case, the lens equation is approximated as
\begin{equation}
\zeta \sim z - {1\over \bar{z}} - {q_p\over \bar{z}-
\bar{z}_{p}},
\label{eq5}
\end{equation}
where $q_p=m_2/m_1$ is the mass ratio between the planet and the host and $z_{p}$ represents 
the position of the planet with respect to the host.  We note that the lengths of $\zeta$ and 
$z$'s in Eq.~(5) are normalized to the angular Einstein radius corresponding to the mass of the 
primary lens, i.e.\ $\theta_{{\rm E},1}=\theta_{\rm E}/(1+q_p)^{1/2}\sim (1-q_p/2)\theta_{\rm E}$.  
For a planetary lens, however, the mass of the primary dominates, i.e.\ $q_p\ll 1$, and thus 
$\theta_{{\rm E},1}\sim \theta_{\rm E}$.  The planet induces two types of caustics, where one 
is located close to the host (central caustic) and the other is away from the host. The central 
caustic has an arrowhead shape and its size is related to the star-planet separation $s_p=|z_{p}|$  
and the mass ratio by \citep{Chung2005}
\begin{equation}
\Delta x_{\rm p}={4 q_p \over 
(s_{p}-s_{p}^{-1})^2
}.
\label{eq6}
\end{equation}
Due to the location of the central caustic close to the host, perturbations induced by the 
central caustic of a planet always appear near the peak of the lensing light curve produced 
by the host of the planet \citep{Griest1998}.

% Figure 2 ------------------------------------------------------
\begin{figure*}[ht]
\epsscale{1.10}
\plotone{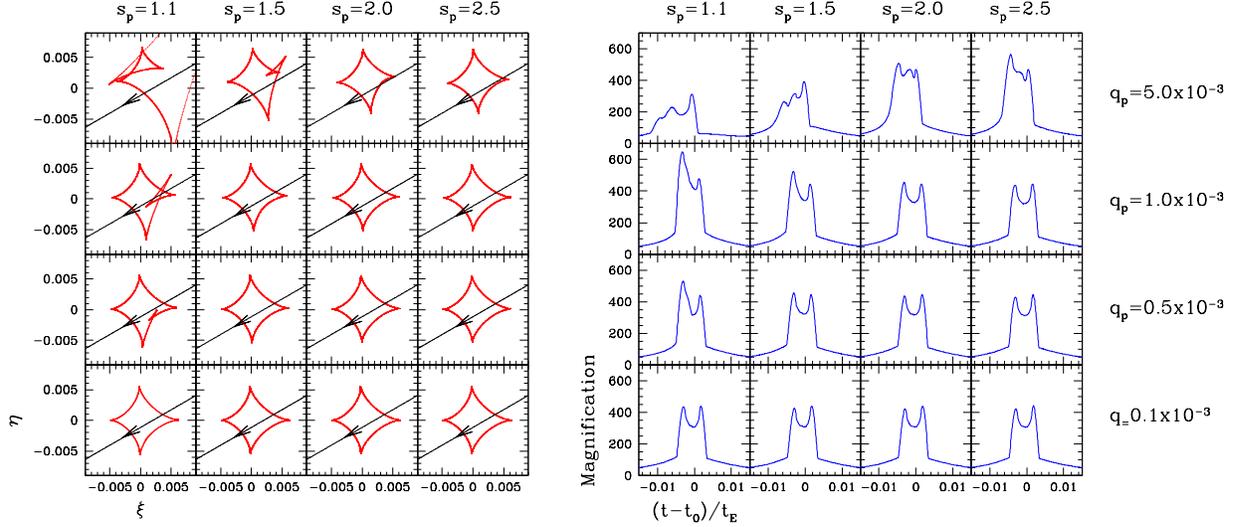}
\caption{Variation of the C-R lensing caustic (left panel) and the light curve (right panel) 
by the presence of planets with various separations $s_p$ and the mass ratios $q_p$. The C-R 
lensing caustic is produced by a binary companion with $s_b=10$ and $q_b=0.5$.  The separations 
$s_p$ and $s_b$ are normalized to the angular Einstein radius corresponding to the total lens 
mass, $\theta_{\rm E}$.  The light curve in each subpanel results from the source trajectory 
(line with an arrow) marked in the corresponding subpanel showing the caustic.  The notations 
$(\xi,\eta)$ represent the coordinates on the source plane where $\xi$ is aligned with the 
binary-lens axis and lengths are normalized to $\theta_{\rm E}$.  The planet is located at 
$(x_p,y_p)=(x_1+s_p\cos\phi,y_1+s_p\sin\phi)$, where $(x_1,y_1)$ is the position of the primary 
lens (heavier binary component) and $\phi=60^\circ$ is orientation angle of the planet with 
respect to the binary axis.
}
\label{fig:two}
\end{figure*}
% --------------------------------------------------------------

The lens equation of C-R lensing is expressed as  
\begin{equation}
\zeta = z - {1\over \bar{z}} + \gamma\bar{z},
\label{eq7}
\end{equation}
which describes the lensing behavior around the primary lens with an external shear $\gamma$.
Here lengths are given in units of the Einstein radius corresponding to the primary mass.
The shear induces a caustic around the lens.  The shape of the caustic is similar to the 
central caustic of a very wide binary lens and thus the C-R lensing provides a good approximation 
in describing the binary lensing behavior in the region around the caustic.  For a very close 
binary lens, the lensing behavior around the center of mass is described by quadrupole lensing 
\citep{Dominik1999}, which also induces an astroidal caustic similar to the C-R lensing caustic.  
To the first order approximation, the caustic size in units of the angular Einstein radius of 
the binary-lens mass, $\thetae$, is related to the separation and the mass ratio between the 
binary-lens components by 
\begin{equation}
\Delta x_{\rm C-R} = {4\gamma \over \sqrt{1-\gamma}};\qquad
\gamma=
{q_b \over s_b^{2}(1+q_b) }, 
\label{eq8}
\end{equation}
for a wide binary and
\begin{equation}
\Delta x_{\rm quad}=4Q \left( 1+{9\over 2}Q \right);\qquad
Q={s_b^2 q_b\over (1+q_b)^2}
\label{eq9}
\end{equation}
for a close binary.

A C-R lensing event can provide a channel to detect planets in binary systems. This is because 
both of the C-R lensing caustic and the central caustic induced by the planet occur in the same 
region around the primary lens and the size of the planet-induced central caustic can be 
comparable to the size of the C-R lensing caustic \citep{Lee2008}. In Figure~\ref{fig:one}, 
we present the sizes of the C-R lensing caustic and the planet-induced central caustic as a 
function of the primary-companion separation. We note that a pair of caustics with separations 
$s$ and $s^{-1}$ have the same size to a linear order and thus we present distributions for only 
wide binaries. The plot shows that the central caustic induced by planets located in the ``lensing 
zone''\footnote{The lensing zone represents the range of the planet-host separations where the 
probability of detecting the planet is high \citep{Gould1992, Griest1998}. The range is 
approximately $1/2 \lesssim s \lesssim 2$, although the range varies depending on how the 
lensing zone is defined.} is of considerable size compared to the size of the C-R lensing 
caustic. This suggests that the C-R lensing anomaly can be additionally affected by the 
planetary perturbation, enabling one to identify the presence of a planet in the binary system. 
In Figure~\ref{fig:two}, we present the variation of the C-R lensing caustic and the lensing 
light curve affected by the presence of planets with various separations and mass ratios between 
the planet and the primary of the binary.  See \citet{Luhn2016} for more details about the 
caustic variation.

We note that the C-R lensing channel enables detections of planets not only in close binaries 
but also in wide binaries. To be dynamically stable, a planet should be either in a circumbinary 
(or P-type) orbit, where the planet orbits the barycenter of the two stars of a close binary, 
or in a circumprimary (S-type) orbit, where the planet orbits just one star of a wide binary 
system. This condition of the planet existence in the binary system matches the lens system 
configuration of the proposed C-R lensing channel of planet detections. Planets in binary systems 
can be detected by various methods such as transit \citep[e.g.,][]{Doyle2011}, eclipsing binary 
timing \citep[e.g.,][]{Qian2010}, and radial-velocity methods \citep[e.g.,][]{Correia2005}. Due 
to the intrinsic nature of the methods, however, it is difficult to detect planets in circumprimary 
orbits and thus all planets in binaries detected by these methods, 19 in total, reside in 
circumbinary orbits.  On the other hand, C-R lensing events can be produced by both close and 
wide binary systems and thus the proposed C-R lensing channel provides a unique channel to 
detect planets both in circumbinary and circumprimary orbits.\footnote{We note that there 
exist two known planets in wide binary systems.  These planets, OGLE-2013-BLG-0341LBb 
\citep{Gould2014} and OGLE-2008-BLG-092LAb \citep{Poleski2014}, were detected by using the 
microlensing method. However, they were detected not through the C-R lensing channel but 
through a repeating channel where the perturbations induced by the planet and the binary 
companion were separately detected.}

\section{Planet Detection Efficiency}

In this section, we demonstrate the high efficiency of the proposed C-R lensing channel in 
detecting planets of binary systems. Estimating the efficiency requires to consider various 
details of observational conditions such as photometric precision and cadence. In order to 
reflect realistic observational conditions, we estimate the detection efficiency for an 
example C-R lensing event that was actually observed by lensing experiments.

% Figure 3 ------------------------------------------------------
\begin{figure}
\includegraphics[width=\columnwidth]{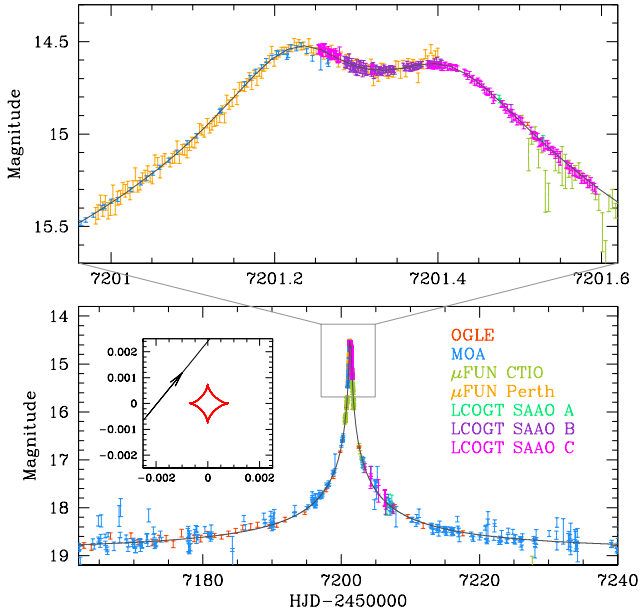}
\caption{Light curve of the microlensing event OGLE-2015-BLG-1319. The lower and upper 
panels show the whole view of the light curve and the zoom around the peak. The inset 
in the lower panel shows the source trajectory with respect to the caustic. The label 
in the legend denote the lensing experiments and the telescopes used for observations.}
\label{fig:three}
\end{figure}
% --------------------------------------------------------------

The event used for our efficiency estimation is OGLE-2015-BLG-1319. The event was analyzed 
in detail by \citet{Shvartzvald2016} and turned out to be an exemplary C-R lensing event 
caused either by a close or a wide binary lens. Figure~\ref{fig:three} shows the light curve 
of the event reproduced based on the same data sets as those used in the previous analysis. 
The model light curve superposed on the data points is obtained from binary-lensing modeling 
based on one of the 8 degenerate solutions (``++ wide'' solution) presented in 
\citet{Shvartzvald2016}. The inset in the lower panel shows the source trajectory with respect 
to the caustic, which has a very characteristic shape of a C-R lensing caustic. The anomaly 
caused by the C-R lensing caustic in the peak of the light curve was densely and precisely 
observed. This was possible because the event was predicted to be a high-magnification event 
before it reached the peak based on the light curve obtained from the survey observation 
conducted by the Optical Gravitational Lens Experiment \citep[OGLE:][]{Udalski2003} and the 
Microlensing Observations in Astrophysics \citep[MOA:][]{Bond2001, Sumi2003}, enabling 
intensive observations of the peak by follow-up observation groups including the Microlensing 
Follow-Up Network \citep[$\mu$FUN:][]{Gould2006} and the RoboNet \citep{Tsapras2009} groups.  
We note that the event was also observed in space using two space telescopes {\it Spitzer} 
and {\it Swift}.  This enabled the determinations of the lens mass and the distance, but we 
do not use the space-based data because these data are irrelevant to our scientific purpose.  
The values of the binary separation and the mass ratio presented in 
\citet{Shvartzvald2016} are $(s_b,q_b)\sim (0.08,0.07)$ for the close binary solution and
and $(s_b,q_b)\sim (14,0.09)$ for the wide solution.  For other lensing parameters,
see Table 1 of \citet{Shvartzvald2016}.

To show the constraint on the existence of a planet, we construct an ``exclusion diagram'', 
which shows the probability of excluding the existence of a planet as a function of the 
planet separation and the mass ratio \citep{Albrow2000, Gaudi2000, Kubas2008, Gould2010, 
Cassan2012}. We construct the exclusion diagram following the procedure of \citet{Shin2015}. 
In this procedure, we first introduce a planet with the parameters $(s_p,q_p,\psi)$ to the 
binary lens with the parameters $(s_b,q_b)$. Here $(s_b,q_b)$ represent the separation and 
the mass ratio between the binary components, while $(s_p,q_p)$ are the separation and the 
mass ratio between the primary of the binary and the planet. The angle $\psi$ denotes the 
orientation angle of the planet with respect to the binary axis connecting the binary lens components. 
We use $(s_b,q_b)$ that are determined from the binary-lensing modeling. For a given set of 
$(s_p,q_p,\psi)$, we then search for other parameters that yield the best fit to the observed 
light curve and compute $\chi^2$ of the fit. We repeat this process for many different 
orientation angles. Then, the probability of excluding a planet for a given $(s_p,q_p)$ is 
estimated as the fraction of the angles $\psi$ that result in fits with 
$\Delta\chi^2 > \Delta\chi^2_{\rm th}$, where $\Delta\chi^2$ is the difference between the 
triple-lens (i.e.\ binary + planet) and the binary-lens models. As a criteria for the planet 
detection, we adopt a threshold value of $\Delta\chi^2_{\rm th}=500$, which is a generally 
agreed value for planets detected through the high-magnification channel \citep{Gould2010}. 
The probability of excluding a planet corresponds to the probability of detecting the 
planet, i.e.\ planet detection efficiency.

% Figure 4 ------------------------------------------------------
\begin{figure}
\includegraphics[width=\columnwidth]{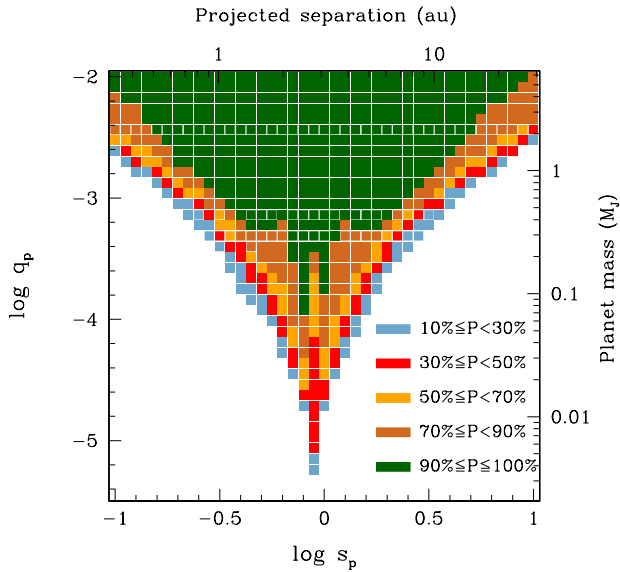}
\caption{Efficiency of detecting planets in the binary lens system responsible for the
lensing event OGLE-2015-BLG-1319 as a function of the normalized separation $s_p$ and the 
mass ratio $q_p$ between the planet and the primary of the binary lens.  The values marked 
in the upper $x$ axis and the right of $y$ axis represent the physical primary-planet separation 
in au and the mass of the planet in Jupiter masses, respectively.  The color coding represents 
the regions of different efficiencies that are marked in the legend.  The planetary separation 
is expressed in units of the angular Einstein radius corresponding to the binary lens mass. 
}
\label{fig:four}
\end{figure}
% --------------------------------------------------------------

Figure~\ref{fig:four} shows the constructed exclusion diagram which shows the planet 
detection efficiency as a function of the normalized separation $s_p$ and the mass 
ratio $q_p$ of the planet.  Since the microlens parallax of the event OGLE-2015-BLG-1319 
was determined using the combined data taken from the ground and space, the physical 
sizes of the primary-planet separation and the mass of the planet were determined.  
By adopting the lens mass of $M_{\rm tot}=0.62\ M_\odot$, the angular 
Einstein radius of $\thetae=0.63$\ mas, and the distance to the 
lens of $D_{\rm L}=4.93$\ kpc, that were determined by 
\citet{Shvartzvald2016}, i.e.\ (+,+) wide model, we convert 
$s_p$ and $q_p$ into the physical sizes of the 
primary-planet separation in au, i.e.\ $a_\perp=s_pD_{\rm L}\thetae$, and the mass 
of the planet in Jupiter masses ($M_J$), i.e.\ $M_p=qM$, and they are presented in 
the upper $x$ axis and the right of the $y$ axis, respectively.

% Table 1 ---------------------------------------------------------------------------
\begin{deluxetable}{lccc}
\tablecaption{Range of planet detection\label{table:one}}
\tablewidth{0pt}
\tablehead{
\multicolumn{1}{c}{Planet}        &
\multicolumn{3}{c}{Efficiency}   \\
\multicolumn{1}{c}{type}          &
\multicolumn{1}{c}{$> 90\%$}      &
\multicolumn{1}{c}{$> 50\%$}      &
\multicolumn{1}{c}{$> 10\%$}       
}
\startdata
Jupiter       & 1.0 -- 10  au   & 0.5 -- 11   au  & 0.4 -- 18 au \\
Saturn        & 2.0 -- 5   au   & 1.0 --  7   au  & 0.9 -- 9 au \\
Neptune       &  --             & 2.0 --  3.5 au  & 1.8 -- 4.0 au    
\enddata                                                                                              
%\tablecomments{}
\end{deluxetable}                                                                                    
%----------------------------------------------------------------------------------------

In Table~\ref{table:one}, we present the ranges of planet detection for 3 different 
planets with masses corresponding to those of the Jupiter, Saturn ($\sim 0.3\ M_J$), 
and Neptune ($\sim 0.055\ M_J$) of the Solar system.  
It is found that the detection 
efficiency is greater than $90\%$ for the Jupiter- and Saturn-mass planets 
located in the ranges of 1.0 -- 10 au and 2.0 -- 5 au from the host, 
respectively.  The ranges with $>10\%$ probability are 0.4 -- 18 au, 0.9 -- 9 au, and 
1.8 -- 4 au for the Jupiter-, Saturn-, and Neptune-mass planets, respecively.  
These ranges encompass wide regions around 
snow lines where giants are believed to form.

Despite the high efficiency of C-R lensing events in detecting planets of binary systems, 
there exists no report of planet detection yet.  The reasons for this can be (1) the 
rarity of C-R lensing events, (2) the unoptimized observational strategy for planet 
detections, and (3) the rarity of planets in the region of sensitivity.  Considering 
the existence of the already known microlensing planets detected through the 
repeating-event channel and those detected by other methods combined with the wide 
region of planet sensitivity of the C-R lensing channel, the reason (3) is unlikely to 
be the main reason of nondetection.  To check the possibilities (1) and (2),  we 
investigate the lensing events reported by the OGLE and MOA surveys in 2015 season.  
From the systematic analyses of all high-magnification events with anomalies near 
the peaks based on the online data of the surveys, we find 12 C-R lensing events 
including  OGLE-2015-BLG-1319, MOA-2015-BLG-040/OGLE-2015-BLG-0318, 
OGLE-2015-BLG-0697/MOA-2015-BLG-148, OGLE-2015-BLG-0797, OGLE-2015-BLG-0812, 
OGLE-2015-BLG-0189 MOA-2015-BLG-085/OGLE-2015-BLG-0472, OGLE-2015-BLG-0313/MOA-2015-BLG-047, 
OGLE-2015-BLG-0033/MOA-2015-BLG-017, MOA-2015-BLG-047, OGLE-2015-BLG-0919, and 
OGLE-2015-BLG-0863.  This indicates that the rarity of C-R lensing events is not the 
reason for the nondetection, either.  However, we find that the coverage of the peak regions 
for all of the C-R lensing events except OGLE-2015-BLG-1319 was not dense enough to 
constrain the existence of a planet.  Therefore, it is likely that the main reason 
for the nondetection of planet is due to the unoptimized observational strategy.  
In other words, this suggests that planets can be detected through the proposed channel 
in abundance with an aggressive strategy to densely cover the peak regions of high-magnification 
events, e.g., vigilant monitoring of high-magnification events and timely alerts of anomalies 
followed by prompt and intensive coverage of the anomalies by follow-up observations

\section{Summary and Conclusion}

We pointed out that C-R lensing events could provide one with an important channel to detect 
planets in binary systems.  We also pointed out that  while other methods could detect planets 
only in circumbinary orbits the proposed C-R lensing channel could provide a unique channel 
to detect planets both in circumbinary and circumprimary orbits.  We demonstrated the high 
sensitivity of the C-R lensing channel to planets in a wide range of planet parameter space 
by presenting the exclusion diagram for an actually observed C-R lensing event.  We mentioned 
that an increased number of microlensing planets in binary systems could be detected with an 
aggressive strategy to densely cover the peak regions of high-magnification events.  The 
sample of an increased number of microlensing planets in binaries will make it possible to 
provide important observational constraints that can give shape to the details of the 
formation scenario which has been restricted to the case of single stars.

\begin{acknowledgments}
Work by C.~Han was supported by the Creative Research Initiative Program (2009-0081561) of 
National Research Foundation of Korea.  
% KREONET
We acknowledge the high-speed internet service (KREONET)
provided by Korea Institute of Science and Technology Information (KISTI).

\end{acknowledgments}


\begin{thebibliography}{}

\bibitem[Abt(1983)]{Abt1983} Abt, H.~A.\ 1983, \araa, 21, 343	
\bibitem[Albrowet al.(2000)]{Albrow2000} Albrow, M.~D., Beaulieu, J.-P., Caldwell, J.~A.~R., et al.\ 2000, \apj, 535, 176
\bibitem[Bond et al.(2001)]{Bond2001} Bond, I.~A., Abe, F., Dodd, R.~J., et al.\ 2001, \mnras, 327, 86
\bibitem[Boss(2012)]{Boss2012} Boss, A.~P.\ 2012, \mnras, 419, 1930
\bibitem[Cameron(1978)]{Cameron1978} Cameron, A.~G.~W.\ 1978, The Moon and the Planets, 18, 5
\bibitem[Cassan et al.(2012)]{Cassan2012} Cassan, A., Kubas, D., Beaulieu, J.-P., et al.\ 2012, Nature, 481, 167
\bibitem[Chang \& Refsdal(1979)]{Chang1979} Chang, K., \& Refsdal, S.\ 1979, Nature, 282, 561
\bibitem[Chang \& Refsdal(1984)]{Chang1984} Chang, K., \& Refsdal, S.\ 1984, Nature, 132, 168
\bibitem[Chung et al.(2005)]{Chung2005} Chung, S.-J., Han, C., Park, B.-G., et al.\ 2005, \apj, 630, 535
\bibitem[Correia et al.(2005)]{Correia2005} Correia, A.~C.~M., Udry, S., Mayor, M., Laskar, J., Naef, D., Pepe, F., Queloz, D., \& Santos, N.\ 2005, \aap, 440, 751
\bibitem[Dan\v{e}k \& Heyrovsk\'{y}(2015)]{Danek2015} Dan\v{e}k, K., \& Heyrovsk\'{y}, D.\ 2015, \apj, 806, 99
\bibitem[Dominik(1999)]{Dominik1999} Dominik, M.\ 1999, \aap, 349, 108
\bibitem[Doyle et al.(2011)]{Doyle2011} Doyle, L.~R., Carter, J.~A., Fabrycky, D.~C., et al.\ 2011, Science, 333, 1602
\bibitem[Erdl \& Schneider(1993)]{Erdl1993} Erdl H., \& Schneider P.\ 1993, \aap, 268, 453
\bibitem[Gaudi et al.(1998)]{Gaudi1998} Gaudi, B.~S., Naber, R.~M., \& Sackett, P.~D.\ 1998, \apj, 502, L33
\bibitem[Gaudi \& Sackett(2000)]{Gaudi2000} Gaudi, B.~S., \& Sackett, P.~D.\ 2000, \apj, 528, 5
\bibitem[Goldreich \& Ward(1973)]{Goldreich1973} Goldreich, P., \& Ward, W.\ 1973, \apj, 183, 1051
\bibitem[Gould \& Loeb(1992)]{Gould1992} Gould, A., \& Loeb, A.\ 1992, \apj, 396, 104
\bibitem[Gould et al.(2010)]{Gould2010} Gould, A., Dong, S., Gaudi, B.~S., et al.\ 2010, \apj, 720, 1073
\bibitem[Gould et al.(2006)]{Gould2006} Gould, A., Udalski, A., An, D., et al.\ 2006, \apj, 644, 37
\bibitem[Gould et al.(2014)]{Gould2014} Gould, A., Udalski, A., Shin, I.-G., et al.\ 2014, Science, 345, 46
\bibitem[Griest \& Safizadeh(1998)]{Griest1998} Griest, K., \& Safizadeh, N.\ 1998, \apj, 500, 37
\bibitem[Hayashi et al.(1985)]{Hayashi1985} Hayashi, C., Nakazawa, K., \& Nakagawa Y.\ 1985, in Protostars and Planets II, eds. D. C. Black \& M. S. Matthew (Tucson: Univ.\ Arizona Press), 1100
\bibitem[Kim et al.(2015)]{Kim2015} Kim, S.-L., Lee, C.-U., Park, B.-G., et al.\ 2015, Journal of the Korean Astronomical Society, 49, 37
\bibitem[Kubas et al.(2008)]{Kubas2008} Kubas, D., Cassan, A., Dominik, M., et al.\ 2008, \aap, 483, 317
\bibitem[Kuiper(1951)]{Kuiper1951} Kuiper, G.~P. 1951\ Proc. Natl.\ Acad.\ Sci.\ U.S.A., 37, 1
\bibitem[Lee et al.(2008)]{Lee2008} Lee, D.~W., Lee, C.-U., Park, B.-G., Chung, S.-J., Kim, Y.-S., Kim, H.-I., \& Han, C.\ 2008, \apj, 672, 623
\bibitem[Luhn et al.(2016)]{Luhn2016} Luhn, J.~K., Penny, M.~T., \& Gaudi, B. S.\ 2016, \apj, 827, 61
\bibitem[Paczy\'nski(1986)]{Paczynski1986} Paczy\'nski, B.  1986, \apj, 304, 1
\bibitem[Poleski et al.(2014)]{Poleski2014} Poleski, R., Skowron, J., Udalski, A., et al.\ 2014, \apj, 795, 42
\bibitem[Pollack et al.(1996)]{Pollack1996} Pollack, J.~B., Hubickyj, O., Bodenheimer, P., Lissauer, J.~J., Podolak, M., \& Greenzweig, Y.\ 1996, Icarus, 124, 62
\bibitem[Qian et al.(2010)]{Qian2010} Qian, S.-B., Liu, L., Zhu, L.-Y., Dai, Z.-B., Laj\'{u}s, E.~F., \& Baume, G.~L.\ 2010, \mnras, 401, L34
\bibitem[Raghavan et al.(2010)]{Raghavan2010} Raghavan, D., McAlister, H.~A., Henry, T.~J., et al.\ 2010, \apjs, 190, 1
\bibitem[Rhie(1997)]{Rhie1997} Rhie, S.~H. 1997, \apj, 484, 63
\bibitem[Safronov(1969)]{Safronov1969} Safronov, V.\ 1969, Evolution of the Protoplanetary Cloud and Formation of the Earth and Planets (Moscow: Nauka)
\bibitem[Shin et al.(2015)]{Shin2015} Shin, I.-G., Han, C., Choi, J.-Y., Hwang, K.-H., Jung, Y.-K., \& Park, H.\ 2015, \apj, 802,108
\bibitem[Shvartzvald et al.(2016)]{Shvartzvald2016} Shvartzvald, Y., Li, Z., Udalski, A., et al.\ 2016, \apj, submitted
\bibitem[Sumi et al.(2003)]{Sumi2003} Sumi, T., Abe, F., Bond, I.~A., et al.\ 2003, \apj, 591, 204
\bibitem[Szebehely(1980)]{Szebehely1980} Szebehely, V.\ 1980, Celest.\ Mech., 22, 7
\bibitem[Thebault \& Haghighipour(2015)]{Thebault2015} Thebault, P., \& Haghighipour, N.\ 2015, Planetary Exploration and Science: Recent Results and Advances, eds. S.~Jin, N.~Haghighipour, W-H.~Ip (Berlin: Springer), 309
\bibitem[Tsapras et al.(2009)]{Tsapras2009} Tsapras, Y., Street, R., Horne, K., et al.\ 2009, Astron. Nachr., 330, 4
\bibitem[Udalski(2003)]{Udalski2003} Udalski, A.\ 2003, Acta Astron., 53, 291
\bibitem[Udalski et al.(1992)]{Udalski1992} Udalski, A., Szyma\'{n}ski, M., Kalu\.{z}ny, J., Kubiak, M., \& Mateo, M. 1992, Acta Astron., 42, 253


\end{thebibliography}
\end{document}